\documentclass[prd,preprint,nofootinbib,showpacs]{revtex4}

\newcommand{\Z}{\mathbf{Z}_2}
\newcommand{\be}{\begin{equation}}
\newcommand{\ee}{\end{equation}}
\newcommand{\bea}{\begin{eqnarray}}
\newcommand{\eea}{\end{eqnarray}}
\newcommand{\N}{\mathcal{N}}

\newcommand{\del}{\partial}

\newcommand{\tr}{\mathnormal{tr}}

\newcommand{\comment}[1]{}

\begin{document}
\preprint{hep-th/0210254}

\title{3-Form Induced Potentials, Dilaton Stabilization, and Running 
Moduli}

\author{Andrew R. Frey}
\email{frey@vulcan.physics.ucsb.edu}
\affiliation{Department of Physics\\
University of California\\ 
Santa Barbara, CA 93106, USA}

\author{Anupam Mazumdar}
\email{anupamm@hep.physics.mcgill.ca}
\affiliation{CHEP, McGill University\\ 3600 University Road\\ Montr\'eal,
QC, H3A 2T8, Canada}

\pacs{11.25.Mj,98.80.Cq,04.65.+e}

\begin{abstract}
We study the potential induced by imaginary self-dual 3-forms in 
compactifications of string theory and the cosmological evolution 
associated with it. The potential contains exponentials of the 
volume moduli of the compactification, and we demonstrate that the 
exponential form of the potential leads to a power law for the scale 
factor of the universe. This power law does not support accelerated 
expansion. We explain this result in terms of supersymmetry and comment 
on corrections to the potential that could lead to inflation or quintessence.
\end{abstract}

\date{\today}

\maketitle

\section{Introduction}\label{s:intro}

If we believe that string theory (or M-theory) is the fundamental 
description of interactions in our universe, then we are obviously 
forced to place the basic processes of cosmology into a string 
theoretic framework.  Important steps have been made in this 
direction by examining four dimensional supergravity models for 
potentials that could support the early phase of the accelerated 
expansion of the universe, known as inflation, which solves some 
of the outstanding problems of the hot big bang cosmology \cite{Linde:1990nc}.
See, for recent examples, \cite{Kallosh:2001gr,Kallosh:2002wj}. 
Other work has identified string theory models in which D-brane 
physics leads to inflation 
\cite{Kallosh:2002wj,Herdeiro:2001zb,Mazumdar:2001mp,Dasgupta:2002ew,
Burgess:2001fx}. 
At the same time, however, it has proven challenging to incorporate 
cosmological acceleration into string theory backgrounds because they
tend to relax to supersymmetric vacua \cite{Hellerman:2001yi,Fischler:2001yj}
(note also that \cite{Burgess:2001fx} found inflation only in a small
region of moduli space).
In this paper, we ask whether a stringy potential generated by higher 
dimensional magnetic fields can give rise to accelerated expansion.  
We restrict our analysis to the classical potential of supergravity.

We study a class of exact solutions to IIB supergravity that have a 
vacuum state (denoted by superscript $(0)$) with $3$-form magnetic 
fluxes that satisfy a self-duality relation 
\be
\label{selfdual}
\star_6^{(0)} \left( F-C^{(0)} H\right) = e^{-\Phi^{(0)}}H
\ee
on the compact space, which should be Calabi-Yau 
\cite{Grana:2000jj,Gubser:2000vg}.  These vacua were described in 
some detail in \cite{Giddings:2001yu} and in dual versions in 
\cite{Gukov:1999ya,Dasgupta:1999ss,Greene:2000gh,Becker:1996gj,Becker:2002sx}.
The metric is of ``warped product'' form, 
\be
\label{warpedmetric}
ds^{(0)2} = e^A \eta_{\mu\nu} dx^\mu dx^\nu + e^{-A} g_{mn} dy^m
dy^n\ ,
\ee
so these models have the phenomenology of the Randall-Sundrum models
\cite{Randall:1999ee,Randall:1999vf,Mayr:2000zd}. The warp factor depends 
on the position of D3-branes (and orientifold planes) on the compact 
space and also determines the $5$-form field strength.  The condition 
(\ref{selfdual}) gives rise to a potential for many of the light scalars,
including the dilaton generically, which vanishes at the classical minimum 
and furthermore has no preferred compactification volume.  We will be 
interested in the behavior of these systems above the minimum, and the 
$4$D metric will generalize $\eta_{\mu\nu}\to g_{\mu\nu}$.

For simplicity, we will mainly consider the case where the internal
manifold is a $T^6/\Z$ orientifold, as described in
\cite{Verlinde:1999fy,Kachru:2002he,Frey:2002hf} (or in dual forms in
\cite{Argurio:2002gv,Becker:2002sx}). We take the torus coordinates to
have square periodicities, $x^m \simeq x^m +2\pi l_s$, so that the geometric
structure is encoded in the metric. On this torus, the $3$-form components
must satisfy the Dirac quantization conditions
\be
\label{quantum}
H_{mnp}=\frac{1}{2\pi l_s} h_{mnp}\ ,\ \ F_{mnp} = \frac{1}{2\pi l_s} 
f_{mnp}\ , \ \ h_{mnp},f_{mnp}\in \mathbf{Z}.
\ee  
Boundary conditions at the orientifold planes give large Kaluza-Klein 
masses to many fields (including the metric components $g_{\mu m}$, 
for example), and the remaining theory is described by an effective $4$D 
gauged $\N = 4$ supergravity with completely or partially broken supersymmetry
via the superHiggs effect 
\cite{Tsokur:1996gr,Andrianopoli:2002rm,Andrianopoli:2002mf,Andrianopoli:2002aq,D'Auria:2002tc,Ferrara:2002bt}.

In the following section, we discuss the dimensional reduction of the
IIB superstring in toroidal compactifications with self-dual $3$-form 
flux, ignoring the warp factor, paying particular attention to the 
potential for a subset of the light scalars.  Next, in section \ref{s:evol}, 
we find the cosmological evolution driven by our potential based on known 
inflationary models; we find that our potentials do not lead to an 
accelerating universe. Finally, in section \ref{s:comments}, we comment 
on the generalization of our results to more complicated models, compare 
our results to other models that do lead to inflation, and discuss 
corrections to our potential that might or might not lead to inflation.


\section{Dimensional Reduction and Potential}\label{s:dimred}

Here we will review the dimensional reduction of $10$D IIB supergravity
in compactifications with imaginary self-dual $3$-form flux on the internal
manifold.  For simplicity and specificity, we will concentrate on 
the toroidal compactifications of \cite{Kachru:2002he,Frey:2002hf}, 
extending our analysis to more general cases in section \ref{s:comments}.
We will ignore the warp factor, which assumes that the compactification
radius is large compared to the string scale\footnote{Actually, because
the warp factor $A$ scales like $R^{-4}$ \cite{Greene:2000gh,Frey:2002hf}, 
the radius need only be a few times the string scale for our approximation 
to be reasonable.}.

\subsection{Kinetic Terms}\label{ss:kinetic}

We will start with the kinetic terms, mostly following the analysis of
\cite{Frey:2002hf}, using the $\N = 4$ 
$SO(6,22)\times SU(1,1)/SO(6)\times SO(22)\times U(1)$ language because 
we are studying configurations away from the moduli space at the bottom 
of the potential. Our main purpose is to identify the physical interpretation 
of the canonically normalized scalars, so we will skip the algebraic details.

As was shown in \cite{Frey:2002hf}, the moduli must be tensor densities in
order to avoid double trace terms in the action,
\be
\label{moduli}
\gamma^{mn}=\frac{\Delta}{2}e^{-\Phi} g^{mn}\ ,\ \ 
\beta^{mn} = \frac{\Delta}{2\cdot 4!} \epsilon^{mnpqrs} C_{pqrs}\ ,\ \
\Delta\equiv \sqrt{\det g_{mn}}\ ,
\ee
along with the D-brane positions\footnote{If the D-branes are coincident,
the index $I$ labels the adjoint representation of $U(N)$; the kinetic 
terms remain the same \cite{Ferrara:2002bt}.} $\alpha^m_I = X^m_I/2\pi l_s$ 
and the $10$D dilaton-axion. For the purpose of cosmology, we want to work in 
the $4$D Einstein frame (note that this is different than in 
\cite{Frey:2002hf} because we are allowing the dilaton to vary)
\be
\label{einstein}
g^E_{\mu\nu} = \frac{\Delta}{2}e^{-2\Phi}g_{\mu\nu}\ .
\ee
From stringy dualities, it can be seen that the moduli definitions 
(\ref{moduli}) correspond to the geometric moduli $g_{mn}, B_{mn}$ in a 
toroidal heterotic compactification, and the metric (\ref{einstein}) is
the $4$D ``canonical metric'' \cite{Sen:1994fa,Schwarz:1993mg} 
in the heterotic description \cite{Frey:2002hf}\footnote{Strictly speaking,
these are only the heterotic dual variables with vanishing fluxes; see
\cite{Becker:2002sx}.}.

The kinetic action obtained from dimensional reduction of IIB SUGRA and 
the D3-brane action is then
\bea
S_{\rm kin} &=& \frac{M_P^2}{16\pi} \int d^4 x \sqrt{-g_E} \left[ R_E+
\frac{1}{4}\del_\mu\gamma_{mn}\del^\mu\gamma^{mn} +\frac{1}{4}
\gamma_{mp}\gamma_{nq} D_\mu\beta^{mn} D^\mu\beta^{qp}-\frac{1}{2}
\gamma_{mn} \del_\mu \alpha^m_I \del^\mu \alpha^n_I \right.\nonumber\\
&&\left. -\frac{1}{2} \del_\mu \Phi\del^\mu\Phi -\frac{1}{2}e^{2\Phi}
\del_\mu C\del^\mu C\right]\ ,\ \ M_P^2 = \frac{1}{8\pi^2 l_s^2}\ .
\label{kinetic}
\eea
Here, $M_P$ is the Planck mass, and we are using a coset space covariant 
derivative 
\be
\label{covder} D_\mu \beta^{mn} \equiv \del_\mu\beta^{mn}+\frac{1}{2}
\left(\alpha^m_I\del_\mu\alpha^n_I-\alpha^n_I\del_\mu\alpha^m_I\right)
\ee
which arises from the magnetic coupling of the D3-branes to $\beta$; this 
is the dimensionally reduced action for the heterotic theory of
\cite{Maharana:1993my,Sen:1994fa}, as one might expect. In deriving
the action, one needs the identity
\be
\label{trick}
\gamma^{mn}\del_\mu\gamma_{mn} = -\gamma_{mn}\del_\mu\gamma^{mn} 
=6\del_\mu\Phi-4\del_\mu\ln\Delta\ .
\ee

It is easiest to study the cosmology of canonically normalized scalars; so
we will break down the geometric moduli. For simplicity we will consider 
only the factorized case $T^6=(T^2)^3$. We can then parameterize the metric 
on an individual 2-torus (say, the $(4-7)$ torus) as
\be
\label{t2complex}
\gamma^{mn}= e^{2\sigma} \left[ \begin{array}{cc} e^{-\zeta}+e^\zeta d^2 &
-e^\zeta d \\ -e^\zeta d & e^\zeta\end{array}\right]\ . 
\ee
Here, $\sigma$ gives the overall size of the $T^2$, $\zeta$ gives the 
relative length of the two sides, and $d$ controls the angle between the
two directions of periodicity. Then the $\gamma$ kinetic term becomes
\be
\label{gammakin}
S_{\rm kin} = -\frac{M_P^2}{16\pi} \int d^4 x \sqrt{-g_E} \sum_{i=1}^3
\left[ 2\del_\mu\sigma_i\del^\mu\sigma_i +\frac{1}{2}\del_\mu\zeta_i\del^\mu
\zeta_i +\frac{1}{2}\del_\mu d_i \del^\mu d_i \right]\ .
\ee
For canonical normalization, the coefficient of the kinetic terms should
simply be $-1/2$, so a further rescaling is necessary.


\subsection{Potential}\label{ss:potential}

The scalar potential comes from dimensional reduction of the background
$3$-form terms in the IIB action. After converting to our variables, 
the potential for the bulk modes is, in generality,
\be\label{Vgeneral}
V= \frac{M_P^2}{4!\cdot 32\pi}  
\left(\det\gamma_{mn}\right)\gamma^{mq}\gamma^{nr}\gamma^{ps}
\left[e^\Phi (F-CH)_{mnp}(F-CH)_{qrs}+e^{-\Phi}H_{mnp}H_{qrs}\right]
\ee
\comment{&=&-\frac{M_P^2}{4!\cdot 32\pi}\int d^4x\sqrt{-g_E}
\left(\det\gamma_{mn}\right)
\gamma^{mq}\gamma^{nr}\gamma^{ps}e^\Phi G_{mnp}\bar G_{qrs}
\label{Vgeneral}\eea}
along with an additional term that subtracts off the vacuum energy
\footnote{This comes from the D3/O3 tension, which must cancel the 
vacuum potential for string tadpole conditions to be satisfied to 
leading order in $l_s$.}. This potential was derived from dimensional 
reduction in \cite{Giddings:2001yu,Kachru:2002he}, from gauged 
supergravity in \cite{Tsokur:1996gr,D'Auria:2002tc}, and from the 
superpotential of \cite{Gukov:1999ya}. One feature to note in this 
potential is that it always has (at least) three flat directions at 
the minimum, corresponding to the radii of factorization 
$T^6=T^2\times T^2\times T^2$.  Also, the $\beta$ moduli do not 
enter into the potential, although some become Goldstone bosons via 
the super Higgs effect
\cite{Tsokur:1996gr,Frey:2002hf,Andrianopoli:2002rm,Andrianopoli:2002mf}.

For cosmological purposes, we will need to have a more explicit form of
the potential in hand.  Since there are 23 scalars $\gamma^{mn}, \Phi, C$,
writing the full potential for a given set of 3-form fluxes would be 
prohibitively complicated, but we can write down a few simple examples and
focus on the universal aspects.

The simplest case is to take the three $T^2$ to be square, so that the
geometric moduli are $\gamma^{44}=\gamma^{77} = e^{2\sigma_1}$, etc., with 
all others vanishing.  Then, above a vacuum that satisfies (\ref{selfdual}),
we can calculate the potential
\bea
V_{\rm dil}& = &\frac{M_P^4}{4(8\pi)^3} h^2 e^{-2\sum_i\sigma_i}\left[ 
e^{-\Phi^{(0)}}\cosh\left(\Phi-\Phi^{(0)}\right) +\frac{1}{2}e^\Phi
\left( C-C^{(0)}\right)^2 -1 \right]\ ,\label{Vdilax}\\
h^2&=& \frac{1}{6}h_{mnp}h_{qrs}\delta^{mq}\delta^{nr}\delta^{ps}
\label{fluxno}
\eea
This potential was written explicitly in $SU(1,1)$ notation in 
\cite{D'Auria:2002tc} and is valid for any 3-form background. The most
important feature of this potential is that there is a vanishing vacuum 
energy, and, further, the radial moduli $\sigma$ feels a potential only 
when the dilaton-axion system is excited.  Since this is the simplest
potential to write down, it will be our primary focus in section 
\ref{s:evol}.  It is very interesting to note that the cosmology of this 
potential for the dilaton-axion has been discussed earlier in 
\cite{Fre:2002pd,Kallosh:2002wj,Kallosh:2002gf} from SUGRA. Importantly,
though, their models did not include the radial moduli or the 
negative term that subtracts off the cosmological constant.

Adding the complex structure is more complicated and more model-dependent.
The simplest possible case, for example, 
$f_{456}=-h_{789}$, is non generic in that (\ref{selfdual}) is satisfied
at $\Phi - \sum_i\zeta_i = C=d_i=0$, so the $\zeta_i$ give extra
moduli compared to other background fluxes (at the classical level).  
However, we still have $\Phi - \sum_i\zeta_i$ fixed by a $\cosh$ potential
with a polynomial in $C,d_i$:
\bea
V_0 &=& \frac{M_P^4}{4(8\pi)^3} h^2 e^{-2\sum_i\sigma_i}
\left\{\cosh\left(\Phi - \sum_i\zeta_i\right) +\frac{1}{2}e^{\Phi+
\sum_i\zeta_i} 
\left[ C^2-2C d_1 d_2 d_3\right. \right. \nonumber\\
&&\left. \left. +d_1^2 d_2^2 d_3^2 + e^{-2\zeta_3}
d_1^2 d_2^2 +e^{-2\zeta_2}d_1^2 d_3^2 + 
e^{-2\zeta_1}d_2^2 d_3^2\right. \right. \nonumber\\
&&\left. \left. +e^{-2\zeta_2-2\zeta_3} d_1^2 +e^{-2\zeta_1-2\zeta_3} d_2^2
+e^{-2\zeta_1-2\zeta_2} d_3^2 \frac{}{}\right] -1 \right\}\label{v0}
\eea
using again (\ref{fluxno}). It is straightforward but tedious to show 
that this potential is positive definite, and the only extremum is at 
$\Phi - \sum_i\zeta_i = C=d_i=0$.  As this case is nonsupersymmetric, 
quantum mechanical corrections should lift the flat directions.

On the other end of the supersymmetry spectrum are the $\N = 3$ models
of \cite{Frey:2002hf}, which fix the dilaton as well as all the complex
structure.  If we ignore $C,d_i$ (set them to a vanishing vacuum value),
we find a potential
\bea
V_3 &=& \frac{M_P^4}{(8\pi)^3} h^2 e^{-2\sum_i\sigma_i} \left[\frac{}{}
\cosh \left(\Phi-\zeta_1-\zeta_2-\zeta_3\right) 
+\cosh\left(\Phi-\zeta_1 +\zeta_2+\zeta_3\right)\right.\nonumber\\ 
& & \left. +\cosh\left( \Phi+\zeta_1-\zeta_2+\zeta_3\right)+
\cosh\left( \Phi+\zeta_1+\zeta_2-\zeta_3\right)-4\right]\ .\label{v3}
\eea
This again has the same $\cosh$ structure for the dilaton; the only 
difference is a factor of $4$ due to the number of components of flux in
the background.

Including the non-Abelian coupling for the D3-brane scalars $\alpha^m_I$
introduces new terms in the potential (see \cite{Ferrara:2002bt} for
a supersymmetry based approach). In the absence of fluxes and even in 
the ground state, this potential is monotonic and simply forces the 
$\alpha^m_I$ to commute.  Otherwise, the branes pick up a $5$-brane
dipole moment and become non-commuting, as discussed in \cite{Myers:1999ps}.
Writing the brane positions as $U(N)$ matrices, the potential is
\bea
V_{\rm b} &=& 2\pi M_P^4 \left[ 2\pi e^\Phi \gamma_{mp}\gamma_{nq}\ \tr
\left( [\alpha^m ,\alpha^n][\alpha^q, \alpha^p]\right)\frac{}{}
\right.\nonumber\\
&&\left.
+\frac{i}{12} (\det \gamma_{mn})^{1/2} e^{\Phi} \left(e^{-\Phi}h-
\star_6(f-Ch)\right)_{mnp}\tr\left(\alpha^m\alpha^n\alpha^p\right)\right]
\ .\label{Vbrane}
\eea
To illustrate this potential, we take $f_{456}=-h_{789}$ as before, 
set $C=d_i=\zeta_i=0$, and consider $\alpha^{4,5,6}\propto I_N$ and 
$\alpha^{7,8,9}=\rho t^{1,2,3}$ with $t^i$ a representation of $SU(2)$.  
Then
\be
\label{braneexamp}
V_{\rm b} = 2\pi M_P^4 \left[ 16\pi e^\Phi\left(e^{-2\sigma_1-2\sigma_2}+
e^{-2\sigma_1-2\sigma_3} +e^{-2\sigma_2-2\sigma_3}\right)\rho^4
+\frac{h_{789}}{2}e^{-2\sum_i\sigma_i}e^{\Phi}\left(e^{-\Phi}-1\right)
\rho^3\right]\ .
\ee
There are actually more terms in this potential as required by supersymmetry;
these are just the lowest order terms that appear in the D-brane
action given by \cite{Myers:1999ps}.  For example, the underlying $\N = 4$
supersymmetry gives a $\rho^6$ term\footnote{We thank S. Ferrara for
discussions on this point.}, and there is also a $\rho^2$ term from 
gravitational backreaction that has been 
calculated using supersymmetry in one case (see \cite{Polchinski:2000uf}); 
in any event,
there is a local maximum in the $\alpha^m_I$ direction.  Like the bulk
potential, this potential has exponential prefactors from the $\sigma$
moduli, and if the bulk scalars are away from their minimum, there is
the same $\exp [-2\sum_i\sigma_i]$ factor.

The key point to take from this discussion of the potential is the
exponential prefactor that appears in all terms, whether bulk or brane
modes.

\comment{\subsection{Scales and Couplings}\label{ss:couple}

We now discuss the various mass scales associated with
the compactification.  In fact, some of our discussion will merely 
reproduce \cite{DeWolfe:2002nn}, except for the fact that we are considering
the Einstein frame metric (\ref{einstein}), which rescales all the masses.

The masses of the fixed scalars, such as the dilaton, are found by 
expanding the potential to quadratic order at the minimum and 
dividing by the kinetic term normalization.  It turns out that all these
masses  go as
\be\label{m2mod}
m_{\rm mod}^2 \sim \frac{M_P^2 h^2}{128\pi^3} e^{-2\sum_i\sigma_i}
e^{-\Phi^{(0)}}
\propto \frac{M_P^2 h^2}{8\pi v(\mathnormal{het})}
\propto \frac{M_P^2 h^2}{8\pi v^2(\mathnormal{IIB})}\ ,\ee
up to a few factors of 2, depending on the precise fluxes (the $16\pi$ comes
from (\ref{Vdilax}).
Here, $l_s^6 v$ is the volume of the $T^6$ in the appropriate heterotic or IIB
variables.  This is also the scale of supersymmetry breaking.  
Note that it is parametrically lower than the Planck mass
as well as the Kaluza-Klein mass scale,
\be\label{m2kk}
m^2_{\rm KK} \propto 
 \frac{M_P^2 k^2}{8\pi v^{2/3}(\mathnormal{het})}\propto 
\frac{M_P^2 k^2}{8\pi v^{4/3}(\mathnormal{IIB})}\ .\ee
(The Kaluza-Klein momentum is $k$.)

We can also discuss various couplings in the theory.  The gauge couplings 
are different for the bulk and brane fields.  Examining the coupling for
BPS winding strings \cite{Frey:2002hf} in a convention with canonical 
propagators for the vectors, the bulk gauge coupling is 
$g^2\sim v^{-1/3}(\mathnormal{het}) \sim v^{-2/3}(\mathnormal{IIB})$,
which is parametrically weak.
The brane gauge coupling, on the other hand, follows from the 
(non-Abelian) DBI action:
$g^2 \sim e^{\Phi^{(0)}}$.  In most cases with flux, the exponential of the
dilaton is near unity at the vacuum, so this is generically strong coupling.  
Both brane and bulk vectors have scalar $\psi F^2$
and axionic couplings, which are suppressed by the Planck mass.}


\section{Cosmological evolution}\label{s:evol}

In this section we seek the cosmological evolution of the dilaton and 
the moduli fields in a flat $d=4$ dimensional space time background.
However, for the purpose of illustration it is prudent that we consider 
a toy model which illustrates the behavior of the potentials 
$V_{dil}, V_{0}$ and $V_3$ described in the earlier section. 
\begin{equation}
\label{pot1}
V\approx e^{-\sum_{i}\alpha_{i}\sigma_{i}} V(\Phi)\, .
\end{equation}
Let us also assume that the above potential has a global minimum 
$\Phi_0$ determined by $V(\Phi)$. At $\Phi_{0}$ the potential vanishes.
In the above, $\Phi$ mimics the dilaton and $\sigma_{i}$ play the role of 
moduli with various coefficients $\alpha_{i}$ determines the slope 
of the potential. For generality we have assumed that there are $i$ number
of moduli. In our original potential all the slopes are fixed at 
$\alpha_{i}=4\sqrt{\pi}/M_P$ (with normalized scalars), see Eq.~(\ref{v0}).
We will model $V_b$ by slightly different potential.

For the sake of simplicity and generality in Eq.~(\ref{v0}), we do not 
assume any form for $d_{i}$ and $\zeta_{i}$ at the moment. It is 
interesting to note that the potential Eq.~(\ref{pot1}) is quite 
adequate to determine the cosmological evolution if they dominate the 
energy density, which is fixed by the value $V(\Phi)$ in our case. 
Further note that $V(\Phi)\propto (M_{P})^4$~~\footnote{Strictly speaking
potential energy ought to be less than $(M_{P})^4$ in order to make sense of 
field theoretic description of the expanding universe.}. 
Therefore, given generic 
initial conditions for all the moduli $\sigma_{i}\sim M_{P}$ in the 
dimensionally reduced action, we hope that the rolling moduli could lead 
to the expansion of the universe. In order to see this clearly, one must 
obtain the equations of motion for both dilaton and moduli if coupled to 
the gravity in a Robertson-Walker space-time metric with an expansion 
factor $a(t)$, where $t$ represents the physical time. The equations of
motion are in the Einstein frame
\begin{eqnarray}
\label{eqm1}
\ddot \Phi +3H\dot \Phi +e^{-\sum_{i}\alpha_{i}\sigma_{i}}V^{\prime}
(\Phi)=0\,, \\
\label{eqm2}
\ddot \sigma_{i}+3H\dot\sigma_{i}-\alpha_{i}e^{-\sum_{i}\alpha_{i}
\sigma_{i}}V(\Phi)=0\,, \\
\label{eqm3}
H^2=\frac{8\pi}{3M_{\rm P}^2}\left[\frac{1}{2}\dot\Phi^2+\frac{1}{2}\sum_{i}
\dot\sigma_{i}^2 +e^{-\sum_{i}\alpha_{i}\sigma_{i}} V(\Phi)\right]\,.
\end{eqnarray}
The Hubble expansion is given by $\dot a/a$, an overdot denotes derivative
w.r.t physical time and prime denotes differentiation w.r.t $\Phi$.

Note that depending upon the slopes of the fields along their classical 
trajectories the dilaton can roll slowly compared to the moduli, in which
case we might be able to solve the moduli equations exactly\footnote{We
are obviously assuming apriori that the dilaton is moving very slowly which
may or may not be the case. Nevertheless, our scenario shall be able to 
discern some of the aspects of the actual dynamics, such as inflationary or
non-inflationary.}. With this simple assumption we first consider 
Eqs.~(\ref{eqm2},\ref{eqm3}) with 
$\dot \Phi \ll \dot \sigma_{i}$, and $V(\Phi)\sim V_{0}$, the latter
condition is true if the dilaton time varying vev changes slowly.
Much stronger condition can be laid on the kinetic terms for the 
moduli and dilaton if we assume
\begin{equation}
\label{cond0}
\dot \sigma_{i} \gg \frac{M_{P}V^{\prime}(\Phi)}{2\sqrt{2\pi}\alpha_{i}
V(\Phi)}\dot\Phi\,.
\end{equation}
The above equation can be derived from Eqs.~(\ref{eqm1},\ref{eqm2}) by 
assuming $\ddot\Phi \ll 3H\dot\Phi$, $\ddot\sigma_{i}\ll 3H\dot\sigma_{i}$ 
and $\dot\Phi \ll \dot \sigma$, which is equivalent to slow-roll conditions.

Now we are interested in solving the moduli field evolution without imposing
slow roll conditions on them. We argue that there exists an attractor region 
with a power law solution $a(t)\propto t^{p}$, which from Eq.~(\ref{eqm3}), 
dimensionally satisfies 
$H^2 \propto t^{-2} \propto e^{-\sum_{i}\alpha_{i}\sigma_{i}} V(\Phi) $.  
Hence we write  
\begin{eqnarray}   
\label{genevol}   
e^{\alpha_{i} \sigma_{i}} & = & \frac{k_{i}}{t^{c_{i}}} \,, \\    
\label{csj}  
\sum_{i=1}^{n} c_{i} & = & 2 \,,    
\end{eqnarray}    
where $k_{i}$ are dimensional and $c_{i}$ are dimensionless constants    
respectively.    
Eq.(\ref{genevol}), coupled with the equations of motion Eq.~(\ref{eqm2})
results in    
\begin{equation}    
(3p-1) c_{i} = \alpha_{i}^2 V(\Phi) \prod_{k=1}^{n} k_{k} \,,    
\end{equation}    
from which we find, using Eq.(\ref{csj}) and Eq.(\ref{eqm2}):   
\begin{eqnarray}     
\label{gensol2}    
V(\Phi)\prod_{k=1}^{n} k_{k} & = & \frac{2 (3p-1)}{\sum_{i=1}^{n}\alpha_{i}^2}
\,, \nonumber \\   
\left(\frac{c_{i}}{\alpha_{i}} \right)^2 & = & \frac{4 \alpha_{i}^2}    
{\left(\sum_{k=1}^{n} \alpha_{k}^2 \right)^2} \,.     
\end{eqnarray}    
When substituted into Eq.(\ref{eqm3}) with $\dot\Phi \ll \dot \sigma_{i}$,
we obtain the key result without using any slow roll condition 
for the moduli where the exponent of the scale factor $a(t)\propto t^{p}$ 
goes as
\begin{equation}   
\label{slope1}    
p = \frac{16 \pi}{M_{P}^2}\frac{1}{\sum_{j=1}^{n}    
\alpha_{j}^2} \,.    
\end{equation}   
We also note that the scaling solution for the moduli fields can be
found quickly as follows for any two moduli, $\sigma_{i}$ and $\sigma_{k}$:
\begin{equation}    
\label{scal1}    
\left(\frac{\dot{\sigma}_{i}}{\dot{\sigma}_{k}} \right)^2 =     
\left(\frac{\alpha_{i}}{\alpha_{k}}\right)^2 \,.    
\end{equation}     
The above equation ensures the late time attractor behavior for all the 
moduli in our case, which has a similarity to the assisted 
inflation discussed in \cite{Liddle:1998jc,Copeland:1999cs}.
From Eqs.~(\ref{genevol},\ref{csj}), we can also write
\begin{equation}
\sigma_{i}=\sigma_{i}(0)-\frac{c_{i}}{\alpha_{i}}\ln t\,,
\end{equation} 
where $\sigma_{i}(0)$ is a constant depending on the initial conditions.

Inflationary solutions exist provided $p > 1$, which can be attained in our
case only when the slopes $\alpha_{i}$ are small enough, or in other words
the moduli should have sufficiently shallower slope. The power law solution
also applies to any $p$ in the range $0< p< 1$, where the expansion is 
non-inflationary.

Note that so far we have neglected the dynamics of the dilaton. In spite 
of rolling down slowly, $\Phi$ eventually comes down to the bottom of the 
potential. So, the prime question is how fast does it roll down to its 
minimum $\Phi_0$. This will again depend on the exact slope of the 
potential for $V(\Phi)$. Nevertheless, if we demand that the dilaton is 
indeed rolling down slowly such as $\ddot \Phi \ll 3H\dot\Phi$, then we 
can mimic the slow-roll regime for the dilaton, and the situation mimics 
that of soft-inflation studied in 
Refs.~\cite{Berkin:1990ju,Berkin:1991nm,Mazumdar:1999tk}. 
\begin{eqnarray}
f(\Phi) &=& f(\Phi_{0})-p~\ln t\,,
\end{eqnarray}
where 
\begin{equation}
f(\Phi)\equiv \frac{8\pi}{M_{P}^2}\int d\Phi \frac{V(\Phi)}{V^{\prime}(\Phi)}
\,.
\end{equation}
Here the subscript $0$ indicates the initial value. 

With $a\propto t^{p}$ and  
$e^{-\alpha_{i}\sigma_{i}}V(\Phi) \propto H^2$, we can then parameterize
the dilaton equation of motion by 
\begin{equation}
\ddot \Phi + 3H\dot\Phi =-cH^2\Phi\,,
\end{equation}
where $c$ is a constant factor which determines the unknown shape parameter
of $V(\Phi)$, which ought to be smaller than one in order to be consistent
with the Hubble equation Eq.~(\ref{eqm3}). In this case, we can find the
exact solution for the dilaton
\begin{equation}
\label{rest}
\Phi(t) \propto a^{-\eta}\,; ~~\eta =\frac{1}{2}\left[\left(3-
\frac{1}{p}\right)-\sqrt{\left(3-\frac{1}{p}\right)^2 -4c}~\right]\,.
\end{equation}

Unlike the dilaton, the moduli have no minimum, and they face the usual 
run away moduli problem. Note that once dilaton reaches its minimum the 
potential Eq.~(\ref{pot1}) vanishes, and so the effective potential for
the moduli. However, once the expansion of the universe driven by
the dynamics for the moduli comes to an end, the dilaton settles down at
$\Phi_{0}$, then the moduli still continue to evolve accordingly 
\begin{equation}
\frac{d}{dt}(\dot\sigma_{i}a(t)^3)=0\,,
\end{equation}
provided there is some source of energy-momentum tensor supporting
the expansion of the universe.  The moduli can indeed come to rest at some
finite value. 

So far we have been concentrating upon the toy model with the potential
Eq.~(\ref{pot1}). Nevertheless, the situation remains unchanged for the
type of potentials we are interested in, see 
Eqs.~(\ref{Vdilax},\ref{v0},\ref{v3}). Note that the dynamical behavior
of the moduli will remain unchanged, but the dilaton may roll slow or 
fast depending upon the actual slope of the dilaton potential.  
By inspecting the potentials we find the corresponding slope 
of the moduli, i.e. $\alpha_{i}=4\sqrt{\pi}/M_{P}$, and $n=3$. 
Therefore, the moduli driven expansion of the universe leads to 
\begin{equation}
p =\frac{1}{3} < 1\,; ~~~a(t) \propto t^{1/3}\,.
\end{equation}
The expansion is non-inflationary and will not solve any of the outstanding
problems of the big bang cosmology. Nevertheless, this expansion which is 
slower than either radiation dominated or matter dominated epoch could be 
the precursor or end stage of inflation in this particular model. 

Now, we briefly comment on bulk potential derived in Eq.~(\ref{braneexamp}).
Note, even if the dilaton is settled down the minimum with $e^{-\Phi}=1$, 
the moduli fields still contribute to the potential. It would then be 
interesting to note whether we get any expansion of the universe from the
moduli driven potential. Further note that the structure of the potential 
is quite different from Eq.~(\ref{pot1}). The potential rather follows 
(taking $\rho$ to be slowly rolling and $\rho \ll 1$)
\begin{equation}
\label{pot12}
V_{b}=32\pi^2 M_{P}^4\rho^4
\sum_{s=1}^{n}\exp\left(\sum_{j=1}^{m}\alpha_{sj}
\sigma_{j}\right)\,.
\end{equation}
This kind of potential has also been solved exactly without using
slow-roll conditions \cite{Copeland:1999cs}. Of course with the 
possibility of some of $\alpha_{sj}=0$ for some combination of $s,j$. 
Our case Eq.~(\ref{braneexamp}) exemplifies with $s,j=1,2,3$.  
For Eq.~(\ref{pot12}), again we demand that 
$\exp\left(\sum_{j=1}^{m}\alpha_{sj}\sigma_{j}\right)\propto 1/t^2$.
The late time attractor solution for the moduli fields can be established
with \cite{Copeland:1999cs}
\begin{equation}
\label{fresult}
\left(\frac{\dot\sigma_{j}}{\dot\sigma_{l}}\right)^2=
\left(\frac{\sum_{q=1}^{n}\alpha_{qj} B^{q}}{\sum_{r=1}^{n}\alpha_{rl}B^{r}}
\right)^2\,.
\end{equation}
In the above equation 
$B\equiv\left(\sum_{j=1}^{m}\alpha_{sj}\alpha_{qj}\right)_{COF}^{T}$, 
where $T$ stands for transpose and $COF$ stands for the cofactor, and
$B^{s}\equiv \sum_{q=1}^{n}B_{sq}$ is the sum of elements in row $s$.
The power law solution $a(t)\propto t^{p}$ can be found to be 
\cite{Copeland:1999cs}
\begin{equation}
\label{fresult1}
p=\frac{16\pi}{M_{P}^2}\frac{\sum_{s}^{n}\sum_{q}^{n} B_{sq}}{\rm det~A}\,,
\end{equation}
where $A_{sq}=\sum_{j=1}^{m}\alpha_{sj}\alpha_{qj}$.

Now, we can read $\alpha_{sj}$ from Eq.~(\ref{braneexamp}). After
little calculation with the normalized $\alpha_{sj}$, we obtain the 
value of $p$ from Eq.~(\ref{fresult1})
\begin{equation} 
p =\frac{3}{16}\ll 1\,.
\end{equation}
Again we find that there is no accelerated expansion. The assisted inflation
in all these cases provides expansion but could not be used to solve inflation
or even late time acceleration during the matter dominated era. In all our
examples we found that the moduli trajectories follow the late time attractor
towards the supersymmetric vacuum. Finally, a word upon supersymmetry breaking
in the observable sector, which will induce mass $\sim 1$~TeV to the moduli 
and dilaton in gravity mediation. Unless the moduli amplitude is damped 
considerably, the large amplitude oscillations of the moduli field will 
eventually be a cause for worry (through particle production). The late 
time moduli domination may lead to the infamous moduli problem 
\cite{carlos:1993cc}.


\section{Discussion}\label{s:comments}

In this section, we would like to comment on the conclusion that we 
cannot get power-law inflation (or quintessence) from the $3$-form 
induced potential.  The reason seems related to comments in 
\cite{Hellerman:2001yi,Fischler:2001yj}; exponential potentials consistent
with the constraints of supersymmetry are generically too steep.  Our
results, then, are consistent with a generalization to many fields of 
the work of \cite{Hellerman:2001yi,Fischler:2001yj} that a system cannot
simultaneously relax to a supersymmetric minimum and cause cosmological
acceleration.  Even though the models considered here do not necessarily
preserve supersymmetry, they are all classically of ``no-scale'' structure,
meaning that they all have vanishing cosmological constant and no potential
for the radial moduli.  So even the non-supersymmetric vacua have 
characteristics of supersymmetric cases.  Furthermore, the potential arises
from the supergravity Ward identity \cite{Tsokur:1996gr,D'Auria:2002tc},
which means it suffers from the same kind of constraints imposed by
the arguments of \cite{Hellerman:2001yi,Fischler:2001yj}.  Heuristically,
the vacua of our system give Minkowski space time, which is static, and
there is no way to accelerate into a static state.

This sort of argument based on supersymmetry is readily generalized 
to the Calabi-Yau models with 3-form fluxes that were studied in 
\cite{Giddings:2001yu}.  Indeed, the form of the bulk mode potential
(\ref{Vgeneral}) is identical, although the complex structure decomposition
of the metric will differ from case to case.  The key thing to note is 
that the overall scale of the internal manifold is always a modulus,
as if we set $\sigma_{1,2,3}=\sigma$.  In fact, it works out so that the
exponential prefactor gives the same $a\sim t^{1/3}$ evolution.  
The potential for brane modes should also be similar, at least for small
non-Abelian parts of the brane coordinates.
Considering a more complicated CY compactification is not the route to 
an accelerating universe. Again, this seems to be a feature of the 
broken supersymmetry.

We should contrast this case to other work that does find inflationary 
physics in supergravity. In the 1980s, \cite{Kounnas:1985vh,Molera:1987ks}
found no-scale supergravities with inflation, but they specified the
potential to give slow-roll inflation.  The freedom to insist on inflation
does not exist here. More recently, other gauged supergravities have been 
found that can give at least a give few e-foldings of inflation
\cite{Kallosh:2001gr,Kallosh:2002gf,Kallosh:2002wj,Fre:2002pd}, but these
do not yet have a known embedding in string theory. These gauged 
supergravities are not of the no-scale type and have a cosmological 
constant. Also, 
\cite{Herdeiro:2001zb,Dasgupta:2002ew,Kallosh:2002wj} describe inflation
based on the motion of branes in a warp factor.  In fact, 
\cite{Dasgupta:2002ew,Kallosh:2002wj} use a background very similar to the
one considered here but include the warp factor.  

There is clearly, then, some hope for finding acceleration in 
compactifications with 3-form magnetic fields, and it is possible to think
of other methods than D3-brane motion.  For example, the warp factor
can modify the potential, although it does not seem likely to change the
basic features.  Another possibility is that the small volume region of
moduli space, where supergravity breaks down, 
has a different form of the potential.  It has been argued that some 
IIB compactifications with flux with one $T^2$ shrinking are dual to 
heterotic compactifications with intrinsically stringy monodromies
\cite{Becker:2002sx,Hellerman:2002ax}, so it is conceivable that 
inflation could occur in such a compactification with a decelerating
end stage described by our model.

Finally, there are many possible corrections associated with supersymmetry
breaking.  It is known that there should be stringy corrections to the
potential in nonsupersymmetric cases and that these would break the
no-scale structure, giving the radial modulus mass (at least in the CY
case) \cite{Becker:2002nn}, and there should also be supergravity loop
corrections.  It would be very difficult to compute this potential, but 
it seems likely that the potential could have a local maximum for the 
compactification radius, allowing for inflation.  There are also 
potentials from instanton corrections, given by wrapped Euclidean D3-branes 
\cite{Frey:2002hf}.  Since the instanton action is proportional to the 
volume of the cycle it wraps, it would actually generate a potential like
the exponential of an exponential.  This type of potential could very
possibly be shallow enough to support inflation, although we have not 
investigated this point.

In summary, we have examined the cosmology induced by $3$-form fluxes in
type IIB superstring compactifications and concluded that the classical
bulk action does not lead to inflation or quintessence because the potential
contains exponential factors that are too steep, much as in
\cite{Hellerman:2001yi,Fischler:2001yj}.  However, we have noted 
loopholes in our analysis which could allow accelerating cosmologies.  
We leave the exploration of those loopholes for future work.

\begin{acknowledgments}
The authors are thankful to Cliff Burgess, Ed Copeland, Sergio Ferrara, 
and Joseph Polchinski for helpful discussions and feedback. 
The work of A.F. is supported by National Science Foundation grant 
PHY97-22022.
A.M. is a Cita-National fellow.
\end{acknowledgments}

\bibliographystyle{h-physrev4} \bibliography{fluxcosmo}

\end{document}